**Title**  : Developing and delivering a remote experiment based on the experiential learning framework during COVID-19 pandemic

**Author information**  :

W.D. Kularatne
Department of Electrical and Electronic Engineering, Faculty of Engineering, University of Peradeniya, Peradeniya, Sri Lanka.
kule@ee.pdn.ac.lk

Lasanthika H. Dissawa[*]
Department of Electrical and Electronic Engineering, Faculty of Engineering, University of Peradeniya, Peradeniya, Sri Lanka.
lasanthikadissawa@yahoo.com
ORCID -  0000-0002-0246-6555

T.M.S.S.K. Ekanayake
Department of Education, Faculty of Arts, University of Peradeniya, Peradeniya, Sri Lanka.
syatigammana@yahoo.com

Janaka B. Ekanayake
Department of Electrical and Electronic Engineering, Faculty of Engineering, University of Peradeniya, Peradeniya, Sri Lanka.
ekanayakej@eng.pdn.ac.lk

*corresponding author

**Abstract**  :
The students following Engineering disciplines should not only acquire the conceptual understanding of the concepts but also the processors and attitudes. There are two recognizable learning environments for students, namely, classroom environment and laboratory environment. With the COVID-19 pandemic, both environments merged to online environments, impacting students' development of processes and characteristic attitudes. This paper introduces a theoretical framework based on experiential learning to plan and deliver processes through an online environment. A case study based on the power factor correction experiment was presented. The traditional experiment that runs for 3 hours was broken into smaller tasks such as a pre-lab activity, a simulation exercise, a PowerPoint presentation, a remote laboratory activity, and a final report based on the experiential learning approach. A questionnaire that carries close and open-ended questions were administered to obtain students' reflections about developing the processes through an online-friendly experiential learning approach. The majority of the students like the approach followed and praise for providing them with an opportunity to perform the experiment in a novel way during the COVID-19 situation.

**Declarations**  :

**Funding** - This research did not receive any specific grant from funding agencies in the public, commercial, or not-for-profit sectors.

**Conflicts of interest/Competing interests** - The authors declare that they have no conflict of interest.

**Availability of data and material** - Not applicable.

**Code availability** - Not applicable.

# Developing and Delivering a Remote Experiment based on the Experiential Learning framework during COVID-19 Pandemic

**Abstract**

The students following Engineering disciplines should not only acquire the conceptual understanding of the concepts but also the processors and attitudes. There are two recognizable learning environments for students, namely, classroom environment and laboratory environment. With the COVID-19 pandemic, both environments merged to online environments, impacting students' development of processes and characteristic attitudes. This paper introduces a theoretical framework based on experiential learning to plan and deliver processes through an online environment. A case study based on the power-factor correction experiment was presented. The traditional experiment that runs for 2 hours was broken into smaller tasks such as pre-lab activity, simulation exercise, PowerPoint presentation, remote laboratory activity, and final report based on the experiential learning approach. The delivery of the lab under online mode delivery was presented. Then students' performance was compared before and after the online mode of delivery. It was found that students' performance in average has a distinct improvement. In order to obtain students' reflections about the online experiential learning approach, a questionnaire that carries close and open-ended questions were administered. The majority of the students like the approach followed and praise for providing them with an opportunity to perform the experiment in a novel way during the COVID-19.

## 1. Introduction

Teaching engineering means more than enabling students to acquire knowledge. It is necessary to foster comprehension of a combination of content, processes and characteristic attitudes related to the topic being studied. Content includes abstract concepts, laws, and theories, whereas processes include observation, classification, measurement, inference, prediction, and communication. Characteristic attitudes involve being curious and imaginative and being enthusiastic about asking questions and solving problems. It is important to focus on the development of these three dimensions among students.

In Engineering disciplines, the content is transferred to students in a classroom environment. The only way to grasp the practical knowledge and experiences for processes is through laboratory experiments. Specially, the experience obtained through experimental work is important as students are focused more on solving problems in real situations. Further, the practical experience gained through laboratory experiments helps to improve skills of applying theoretical knowledge in practical situations (L. Feisel et al., 2002).

The COVID 19 pandemic has imposed a global shutdown of various activities, including educational activities. This resulted in transforming the classroom learning environment into an online learning environment. The challenges and the opportunities of crisis-response migration methods of universities are discussed in (Adedoyin & Soykan, 2020). With COVID 19 pandemic, the delivery of the content is done online successfully using platforms like Zoom and Microsoft Teams. Reference (Tang et al., 2020) shows that flipped learning improved students' learning, attention, and evaluation of courses. However, there is always a question among engineering educators, whether they are able to develop processes and characteristic attitudes of students when switched to online learning environment.

Reference (Almeida et al., 2009) state that in order to provide a good context for making students effectively understand what they are being taught, tasks designed as investigation activities and uses real data and/or asking for problem-solving, that is, learning by experience, can effectively be used. In experiential learning, the 'learning' relies on the practical aspects and the experience being considered the key to success in the educational act. This approach based on experience adds good value to the student's individuality, develops his/her action skills, reflection skills, critical and innovative thinking, initiative, motivation, curiosity, and trust in his/her own person (Gorghiu & Santi, 2016).

According to (Kolb, 1984) experiential learning consists of four stages: concrete experience, reflective observation, abstract conceptualization and active experimentation. That is, learning starts with the active involvement of a learner in getting experience by doing something individually or as a team; then the learner taking a time-out from 'doing' and stepping back from the activity and reviewing what has been experienced and done. The above stages are followed by the process of making sense of what has happened and involve interpreting the events and understanding the relationships between them, drawing upon theory from textbooks for framing and explaining events, models they are familiar with, ideas from colleagues, previous observations, or any other knowledge that they have developed. Finally, the learner considers how they are going to put what they have learnt into practice. This cycle is shown in Figure 1. An improved form of Kolb's experiential learning cycle consists of contextually rich concrete experience, critical reflective observation, contextual-specific abstract conceptualization, and pragmatic active experimentation is proposed in (Morris, 2020).

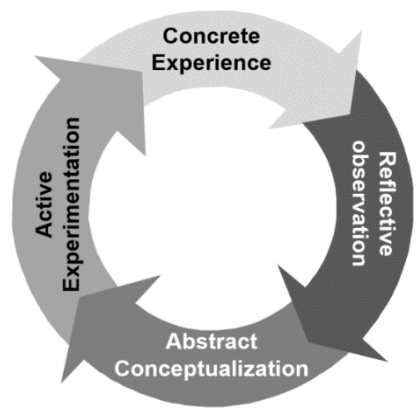

Figure 1 Experiential learning cycle

## 2. Online Modes of Getting Concrete Experience

Prior to the pandemic, most of the experiments were performed in the laboratory environment and simulation environment. In the laboratory environment, the students work on real equipment and instruments, and they obtained the skills to use measuring equipment and experiences of real-world practical problems. On the other hand, the simulation laboratories are conducted as a pre-lab exercise to obtain an idea of the actual outcome of the experiment (L. D. Feisel & Rosa, 2005) before performing it in a real laboratory and also they are conducted for explaining theoretical concepts (Balamuralithara & Woods, 2009). Further, simulation laboratories are conducted as an alternative way of doing experiments related to expensive or large systems, which are not practical for doing in a university laboratory environment. Simulations are also used to demonstrate the events that are not easily seen, such as current flow, heat transfer and electromagnetic fields (Bourne et al., 2005). As the software represents simplified mathematical models of complex real systems (Mosterman et al., 1996), in some cases, the simulation does not provide accurate results.

During the COVID-19 pandemic lock-down period, many universities explored online modes of performing experiments such as (a) simulation labs, (b) remote labs and (c) virtual labs (Odeh et al., 2013).

From simulation-based laboratories, the experience and the practical knowledge obtained by the student depends on the capabilities, authenticity and constraints of the software (Ertugrul, n.d.). But it can be conducted as a substitution for some experiments, as mentioned earlier. Reference (Das, 2018) proposes a MATLAB-Simulink modelling based experiment to understand the characteristic of solar PV cells and solar PV systems. A virtual microgrid experiment introduced in (Chai et al., 2020) uses Simscape, an electrical power system toolbox in MATLAB/Simulink software package, to model the microgrid. Students can download and install software packages and libraries on their personal computers and can do the experiment remotely. The simulation tools give a robust platform to create models and help to analyze the behaviour and performance of systems but do not provide the students with a feeling of the real presence of laboratory equipment (Peterson & Feisel, n.d.).

A laboratory that could give access to operate and control the real equipment via the internet is called a remote lab. This allows the students to undertake the experiments through the internet. Students can access the remote lab using their personal computers via a web browser application and can send commands to control the lab equipment. The commands will go through a server and execute the command in the real equipment. The results of the experiment will be displayed at the student's computer. For example, a web API that conforms to the Web of Things standard to control a microscope was developed (Collins et al., 2021). It provides a modern graphical control interface and allows multiple microscopes to be controlled by one computer. Further, it facilitates sharing of equipment between local or remote users. The steps of implementing remote microscopy are discussed in (Goldberg & Dintzis, 2007). A digital camera that was attached to a light microscope provides the images of slides. The students can control this light microscope from a remote location via virtual microscopy software. The authors (Odeh et al., 2013) created a remote electronic Engineering lab based on Augmented Reality using a video camera and real experimental electronic tool kit. The camera transfers a live video of the electronic kit at the remote lab. In this remote lab, students can draw circuit connections on a webpage. Then, after the verification of the connections, the data were emulated onto a real multimeter. The remote lab provides experiences of practical issues that would not occur in a simulation environment (Ferreira et al., 2002). Therefore, remote labs are good choices for distance learning, and it allows the students to do experiments on real equipment located at a distance through the internet.

The virtual lab is not a real lab, but the entire infrastructure required for a real lab setup is obtained through computer-generated graphics, and it generates results from software simulations (Ma & Nickerson, 2006). Some virtual labs are developed only using computer-generated graphics, and some are developed using computer-generated graphics, virtual reality sensors, and leap motion control device. In the latter, virtual reality sensors will capture the overall body movements and send the information to the computer to render them in the virtual environment. Further, the hand and finger movements are identified by the leap motion device, and it communicates with the virtual reality sensor and transfers this data to the computer to translate the movements of the students into a virtual environment. The graphics in a virtual environment are maintained in the same form as a real device. This kind of virtual laboratory is effective in increasing student's knowledge and understanding of handling equipment as students can visualize and experience the whole experimental process. Reference (Hasan et al., 2020) presented a virtual electric machines laboratory using Oculus Rift, Unity3D, and Leap Motion to do experiments in a safe environment to gain a broad understanding of the concepts of how electric machines work. A virtual instrumentation and measurement laboratory was reported in (Valdez et al., 2014). Since it used 3D components, students can get a real visualization of the circuit components. But this is a full software program (no hardware components were included) that will not provide real results. A virtual lab for real-time control of a mobile robot is presented in (Solak et al., 2020). In this lab setup, an IP camera was fixed to monitor the indoor laboratory and the mobile robot in real-time. The students can place a virtual target or virtual obstacles anywhere on the video generated by the IP camera. The navigation of the robot is monitored through the personal computer web server. The web server on the single-board computer in the robot can communicate with the student's PC, and it can execute the developed application software on the robot. Further, the robot can control manually through the web environment. From this kind of simulation-based virtual labs, students do not gain experience in analyzing and interpreting real-world results.

Even though the literature provides different online approaches that can be applied to develop skills and experience related to processors, they do not provide a theoretical framework that can be used when developing and delivering online experiments. The paper presents an experiential learning approach for a remote power factor correction experiment as a case study.

## 3. Method

According to (Osipov et al., 2015), the ideal online lesson duration is 15–20 min. Its further states that it is hard for both the teacher and the student to study for more than 30 min at a time. Further, reference (Basilaia & Kvavadze, 2020), which reports a transition to the online mode of delivery during the COVID-19 pandemic, states that when the online teaching has started, changes in the duration of the online lessons were done to avoid prolonged contact of the students with a computer. Considering these facts, the usual laboratory session was changed while following the experiential learning cycle shown in Figure 1.

This laboratory session was given to the students after a comprehensive lecture. In order to provide the laboratory session as short duration lessons, it was planned as described in Table 1. A PowerPoint presentation was used to reiterate the subject content after they have completed the first simulation exercise.

Table 1 Activities based on the experiential learning cycle

| Experiential learning cycle | Activity | Description |
|---|---|---|
| **Active experimentation** | Pre-lab work | This is a personalized activity and discusses in more detail in Section 3.1. |
| **Concrete experience and Reflective observation** | Simulation and reflections | A simulation followed by a number of short questions were included for students to reflect on the simulation activity. The details of this activity are given in Section 3.2. |
| **Abstract conceptualization** | PowerPoint presentation | As described in Section 3.3, a PowerPoint presentation was given for students so that they could connect what they learn to theory |
| **Concrete experience** | Remote laboratory activity | They could connect to the laboratory setup remotely and carry out a simple power factor correction experiment. This is described in Section 3.4. |
| **Reflective observation & Abstract conceptualization** | Final report | This is designed for students to reflect on observation and connect it to the field. This is described in Section 3.5. |

### 3.1. Pre-lab Work

This contains a number of short tasks that help students to connect the activities that they do into the real world. All the tasks were based on a single-phase pump load connected to a 230 V rms supply. In order to personalize the activity, the capacity of the pump was tie with the student's registration number. The capacity of the pump is 7.5Z kW (where Z is the reminder of the [Registration number/3] plus 1), and the operating power factor at 100% loading is at 0.87. The tasks are given in Table 2.

Table 2 Tasks given in the pre-lab work

| Task | Description |
|---|---|
| 1 | Represent the load by a resistor (R) in series with an inductor (L) |
| 2 | The pump is connected to a distribution board 20 m apart. Select a suitable cable to supply the pump (cable data was given) |
| 3 | Calculate the capacitance required to improve the power factor to 0.99 |
| 4 | Compare the power loss and voltage drop across the cable without and with the capacitor |

### 3.2. Simulation and Reflections

This simulation exercise is based on the 'Circuit Simulator Applet' available at https://falstad.com/circuit/. The reason for using this Applet is easy accessibility. In the usual classroom, PSCAD is used. The installation of this software will need special support, and also, data charges to download the software is not affordable to some students. This simulation is based on the pre-lab work that they did. An instruction sheet was given to use the Circuit Simulator Applet. The tasks are given in Table 3.

Table 3 Tasks for the simulations and reflections

| Task | Description |
|---|---|
| 1 | Implement the R-L representation of the pump load considered in the prelab in 'Circuit Simulator Applet.' |
| 2 | Using the scopes available in the Applet, obtain the waveforms of load current, load voltage, and power consumed by the resistive part of the load and inductive part of the load. |
| 3 | Using the waveform obtained, calculate the power factor of operation and compare it with the calculated value. |
| 4 | Implement the above load with the power factor correction capacitor in 'Circuit Simulator Applet' and obtain the load voltage and current. |
| 5 | Obtain using the Applet the losses in the cable when the pump is operating at 100% loading without and with the power factor correction capacitor and compare the results with the calculated value |
| 6 | Reflect on the calculated and simulated results for any discrepancies and write down reasons for such discrepancies. |

### 3.3. PowerPoint presentation

This presentation carried different types of real loads and their R-L representation, consequences of low power factor operation, power factor measurement techniques, and power factor correction. Some slides are shown in Figure 2.

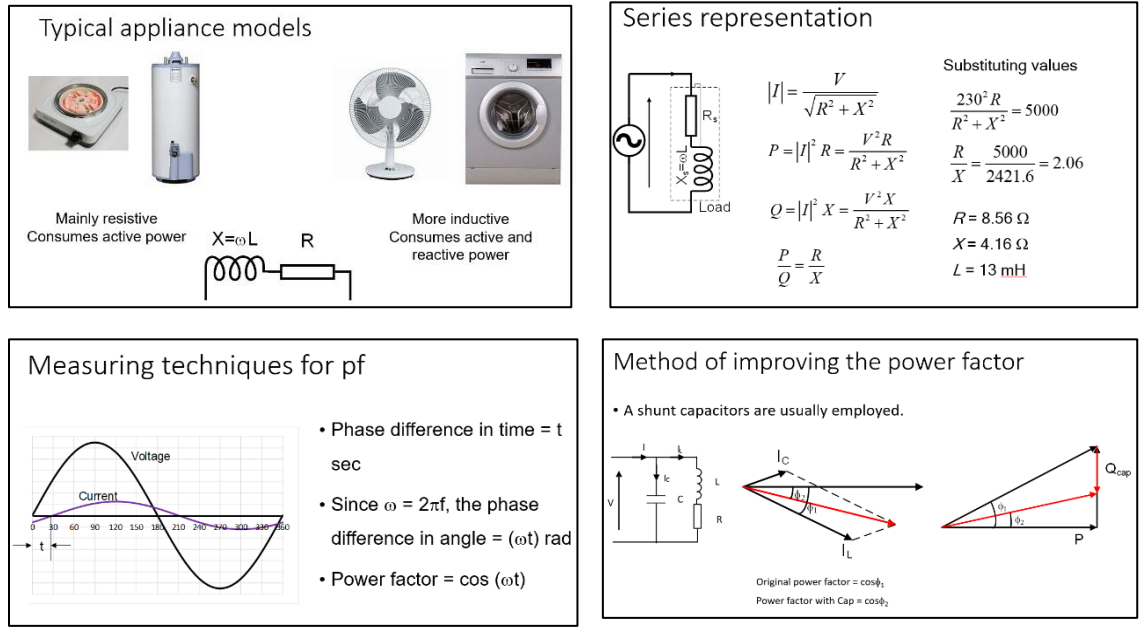

Figure 2 Some slides provided for abstract conceptualization

### 3.4. Online laboratory activity

This section provides comprehensive commentary on the development of the laboratory setup. Figure 3 provides the overall setup developed. The power factor improvement circuit consists of a resistor, inductor, capacitor, relay switch and a variac. The sensor circuit consists of a current sensor, voltage sensor, an op-amp circuit to obtain power factor and Arduino UNO controlling board. Raspberry pi and the oscilloscope were exposed to the internet using two real IPs such that they can be accessed through the internet from any network.

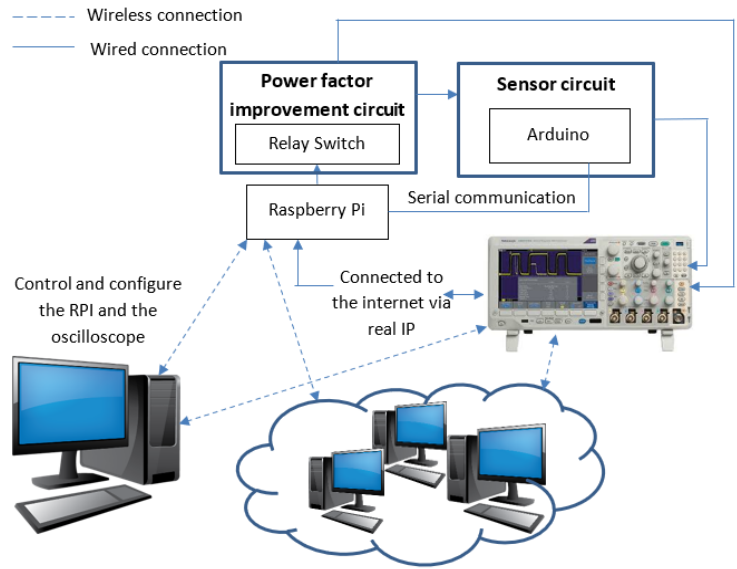

Figure 3 Components of the online laboratory setup

### 3.4.1. Power factor improvement circuit

Figure 4 shows the hardware setup used in the power factor correction experiment. The relay switch is used to connect or disconnect the capacitor to the RL load. The relay signal is given via a raspberry pi digital output port. The relay ON and OFF commands were given to the raspberry pi by a remote user via a web application.

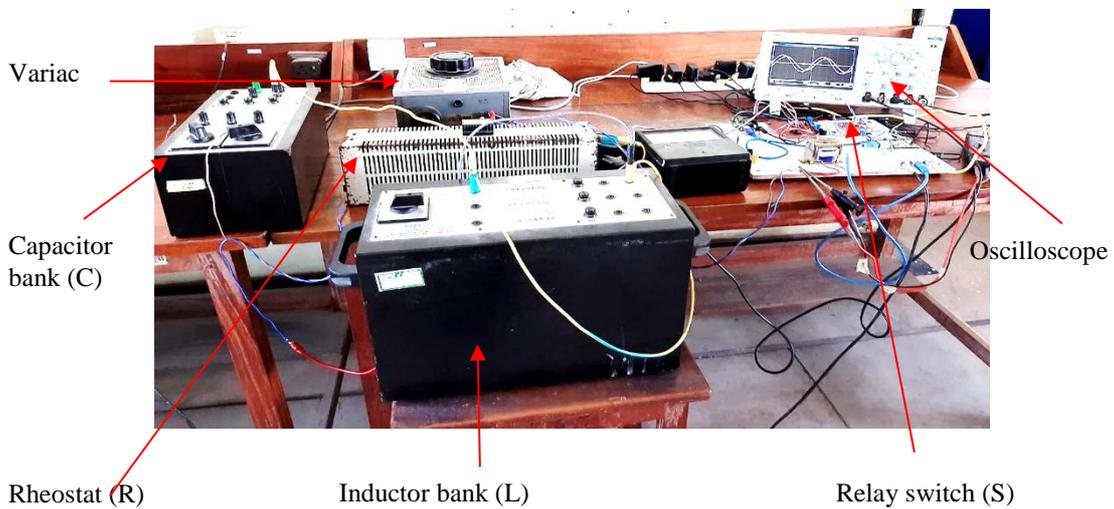

Figure 4 Hardware setup

### 3.4.2. Sensor circuit to obtain voltage, current and power factor measurements

Figure 5 shows a diagram of the sensor circuit. It consists of current and voltage sensing devices, op-amp ICs, an XOR gate IC and Arduino-UNO microcontroller board. The voltage and current sensing devices were used to measure the voltage and current of the circuit. To obtain the power factor, the op-amp comparator circuit was used with an XOR gate. The sensing signals were sent to the analog input pins in the Arduino board, and the pulse generated by the XOR gate IC was also set to a digital input pin of the Arduino board. The mathematical calculations were bone inside the Arduino board to obtain root mean squared voltage, root mean squired current, and the power factor.

### 3.4.3. Web-based Oscilloscope

To observe the circuit current and voltage waveforms, an oscilloscope was used with a current probe and isolated voltage probes. The Oscilloscope, current probe and voltage probes used were Tektronix MDO 3014, Tektronix A622 and GW-Instek GDP-025. The MDO 3014 oscilloscope was used, which can access through the internet. The URL of the oscilloscope was placed as a button in the GUI to enable students to access it.

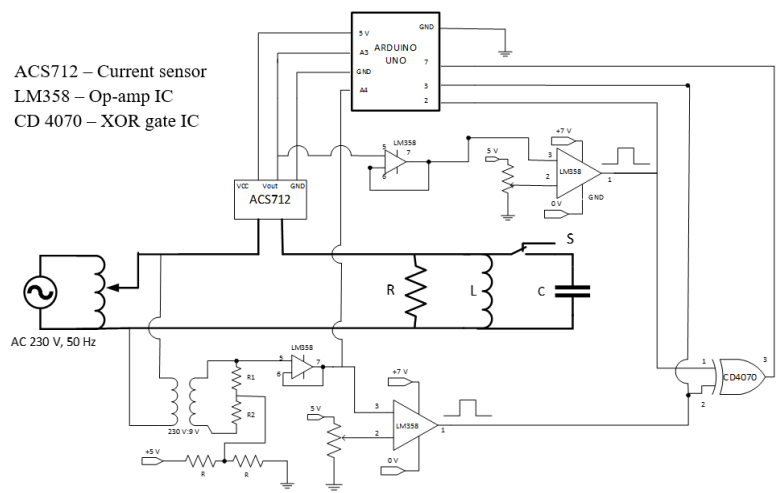

Figure 5 Sensing circuit

### 3.4.4. The graphical user interface of the system

Figure 6 shows the graphical user interface (GUI) of the lab setup. HTML and CSS coding were used on the GUI programming. The interface shows the voltage, current and power factor measurements of the RLC circuit. Further, to operate the relay switch (to add or remove the capacitor from the RLC circuit), a button was placed. The button toggles between 'add capacitor' and 'remove capacitor'. Furthermore, a link was added to access the oscilloscope.

    In order for students to explore a different aspect of the experiment individually, an instruction sheet was developed while providing clear instructions. Each student was assigned a specific time to access the laboratory setup. This was to minimize any issues that may create due to high traffic. The tasks that students were asked to carry out are summaries in Table 4.

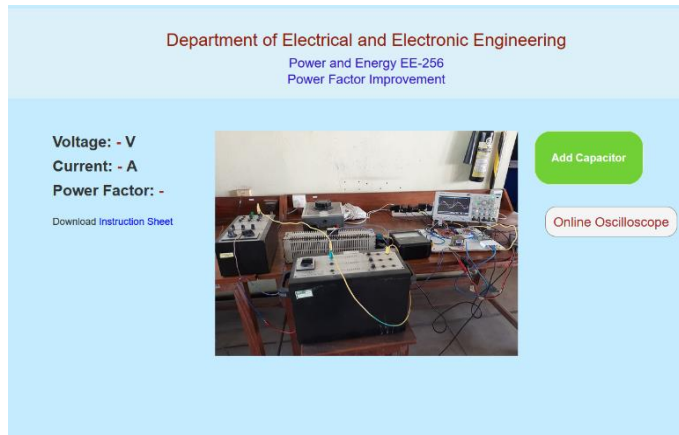

Figure 6 Graphical user interface of the lab setup

Table 4 Activities in the online lab

| Task | Activity |
|---|---|
| 1 | When the toggle switch is at the 'add capacitor' position, obtain the measurements of voltage, current and power factor. |
| 2 | Go to the oscilloscope window and observe the waveform patterns. Also, using the oscilloscope settings, calculate the power factor. |
| 3 | Go back to the initial window |
| 4 | Click on the "Add Capacitor" to on the switch the power factor correction capacitor |
| 5 | Obtain the voltage, current and power factor readings |
| 6 | Go back to the oscilloscope window and observe the waveform patterns. Also, using the oscilloscope settings, calculate the power factor |

*3.5. Final report*

The final report is a formal report where students are asked to report the results of each activity carried out under Sections 3.1. to 3.4. Students are then asked to add a discussion about discrepancies of different results and reasons for them. It is anticipated that this report will be a take-home guide to apply their knowledge to wide applications when they graduated as Engineers.

## 4. Student Performance and Reflections

Figure 7 shows the marks distribution of student who did the experiment on power factor correction from 2017 to 2021. From 2017 to 2019, the experiment was done in the laboratory. Students submitted a pre-lab report as described in Table 1 prior to doing the experiment and then submitted the final report. The main difference between the way the experiment was conducted prior to the COVID period and after 2020 were no simulation activity, a 2-hour long laboratory experiment, and conducting the experiment in groups of 3 students. In 2020, students individually did the first three steps in Table 1, i.e., pre-lab work, simulations and reflections, and PowerPoint presentation. The remote laboratory setup was not available that year. This paper describes the steps followed in 2021. The main difference noted when the student did the experiment as groups and as individuals is existence of some outliers in 2020 and 2021. More importantly, on average, students' performance was improved when adopting the methodology described in this paper.

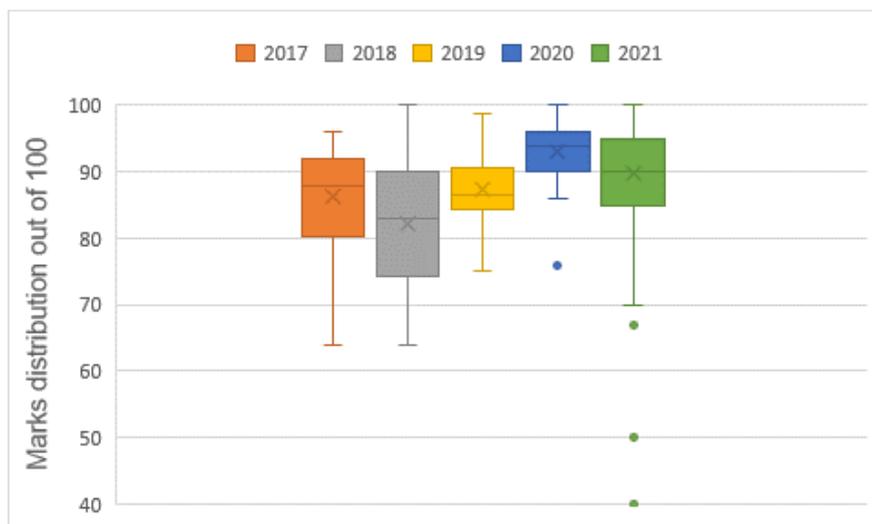

Figure 7 The marks distribution of student who did the experiment on power factor correction from 2017 to 2021

In order to assess whether students fulfil the intended learning outcomes, a google form was developed that provided some close-ended questions and open-ended questions. Table 5 summaries the responses to some of the close-ended questions:

Table 5 Questions and Student's responses

| Question | Strongly Agree % | Agree % | Other % |
|---|---|---|---|
| Narrated PowerPoint presentation provided a good insight into the online experiment | 51.7 | 40 | 8.3 |
| The simulation was intellectually simulating | 16.7 | 46.7 | 36.6 |
| Simulation and PowerPoint provided a good base so that I could connect this experiment to real world | 23.7 | 33.9 | 42.4 |
| The instruction sheet provided for the online lab was informative and useful | 42.4 | 50.8 | 6.8 |
| Allowing to visualize the oscilloscope trace was useful | 20.0 | 46.7 | 33.3 |

For the close-ended question, 'under COVID-19 learning from home situation, the online lab backed by PowerPoint presentation and simulations was an ideal alternative,' the response is shown in Figure 8.

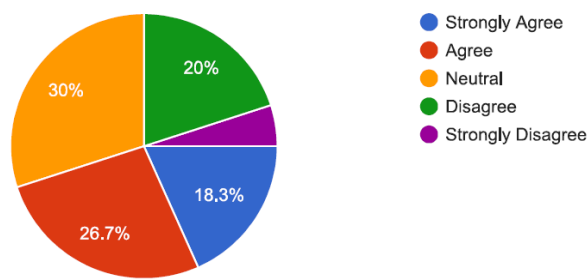

Figure 8 Student responses

Some of the responses to the open-ended questions are given in the following quotes:

*"It is a good opportunity for us to complete our lab assignments online in this pandemic situation. But we are missing the hand on experience, which is really important".*

*"With the situation of the country, this way of conducting lab sessions is very useful as we can come with a good idea about the lab session though we couldn't make our own setup. This gave me a good experience never had before. Allowing time slots to use the setup individually is really good as everyone can engage in this lab session. Everything was well planned and thank you for trying something new".*

*"The instruction sheet and the video before the lab is very understandable and sufficient for doing the lab".*

*"I think it's a very good idea to provide us with a virtual laboratory experience during this period as it can be considered as one of the best alternatives for physical labs. I believe that this can be improved to a great extent by adding more features such as the ability to change component values. In addition, login credentials (or something similar) can be assigned to us so that only one student can access the lab during a given time slot. The key feature I saw in this virtual lab concept was that we are seeing a realistic result, rather than a simulated result".*

However, due to some technical hiccups, some students had issues. They responded as:

*"Interface of virtual lab is good if all functions were working. If all are working well, this approach is ideal to do our lab classes".*

*"When I was trying to do the lab, I could not access the lab in the time slot that was assigned to me. The setup was not working".*

*"The values of voltage, current and the power factor were correctly updating, and the "Add/Remove Capacitor" button was working. The only problem was that we were not able to access the oscilloscope screen due to a connection error. It would have been a really good experience if we were able to connect to it and obtain the necessary readings".*

*"There were some connection issues at the beginning. But later on, everything was fine and successfully finished".*

## 5. Conclusions

It was recognized by many Engineering educators that processes associated with some modules could not be delivered effectively through the online environment. Moving on to the online mode for laboratory experiment during the COVID-19 pandemic was new to many educators, and often they had doubts about the planning and delivery of laboratory experiments. Considering these facts, this paper presented a theoretical framework based on experiential learning to plan and deliver experiments online. Since a laboratory session online should be shorter, the proposed theoretical framework ideally suited to deliver laboratory session in a number of shorter activities. A case study based on the power factor correction was used to demonstrate the planning and execution of the proposed delivery mode. Students' performance was compared before and after the online mode of delivery, and it was found that students' performance was improved when the laboratory activity was conducted as described in this paper. Students' opinion about this mode was obtained using an online questionnaire. In general, students like the idea of a remote lab and the way it was delivered. A few issues with the laboratory setup were also highlighted. It was recognized that a better way of accessing the oscilloscope is important, and latency should be improved.

**Data Availability**
No additional data were used to support this study.
**Conflicts of Interest**
The authors declare that there is no conflict of interest regarding the publication of this paper.
**Funding Statement**
This research was not funded.